\documentstyle[preprint,tighten,aps,amstex,graphicx,times,floats,epsf]{revtex}

\newcommand{\ie}{{\it i.e.}\;}

\providecommand{\SP}{\scriptscriptstyle}

\newcommand{\cp}[1]{\tilde{\chi}_{#1}^+}

\providecommand{\mg}{m_{\tilde{g}}}
\newcommand{\msn}[1]{m_{\tilde{\nu}_{\SP {#1}}}}
\newcommand{\mch}[1]{m_{\tilde{\chi}^+_{\SP {#1}}}}
\newcommand{\mne}[1]{m_{\tilde{\chi}^0_{\SP {#1}}}}
\newcommand{\msb}[1]{m_{\tilde{b}_{\SP {#1}}}}

\newcommand{\lt}{\left}
\newcommand{\rt}{\right}
\newcommand{\ov}{\overline}

\newcommand{\nn}{\nonumber \\}
\newcommand{\no}{\nonumber }
\def\openone{\leavevmode\hbox{\small1\kern-3.8pt\normalsize1}}%

\newcommand{\gev}{\,\mbox{GeV}}
\newcommand{\mev}{\,\mbox{MeV}}
\newcommand{\tev}{\,\mbox{TeV}}

\newcommand{\eq}[1]{eq.~(\ref{#1})}

\newcommand{\met}{{\rm E}\hspace{-0.45em}|\hspace{0.1em}}

\newlength{\nseparation}
\newcommand{\npb}[3]{${\rm Nucl. Phys.}$ {\bf B#1} (#2)~#3}

\renewcommand{\prd}[3]{${\rm Phys. Rev.}$ {\bf D#1} (#2) #3}

\newcommand{\hep}[1]{${\tt hep\!-\!ph/}$ {#1}}

\begin{document}

\thispagestyle{empty}

\title{ 
~\vspace*{-1cm} \\
{\normalsize \rm
\parbox[t]{5cm}{
FERMILAB-Pub-00/211-T\\
MAD--PH--00--1190 }
\hfill hep-ph/0008321  
}\\[5mm]
Probing Light Sbottoms with $\bf B$ Decays
}

\author{  Ulrich Nierste$^1$\footnote{Fermilab is operated by URA 
                  under DOE contract No.~DE-AC02-76CH03000.},
          ~~~~~~~~~Tilman Plehn$^2$\footnote{Supported in part by DOE grant 
                  DE-FG02-95ER-40896 
                  and in part by the University of Wisconsin Research 
                  Committee with funds granted by the Wisconsin Alumni 
                  Research Foundation.} } 

\address{\vspace*{2ex}
$^1$Fermi National Accelerator Laboratory, Batavia, IL 60510, USA\\
$^2$Department of Physics, 
University of Wisconsin, Madison, WI 53706, USA
} 

\maketitle 

\begin{abstract}
  We discuss recently reported experimental hints for a bottom 
  squark with mass around 3.5 $\gev$ decaying as 
  $\tilde{b} \to  c \ell \tilde{\nu}^* $. We correlate the 
  sbottom lifetime with the decay rates for 
  $b \to \tilde{b} \nu \tilde{\nu}{}^* $ and 
  $b \to \tilde{b} \ov{\nu} \tilde{\nu} $ in the framework of 
  a minimal supersymmetric model extended by right-handed 
  (s)neutrinos. Confronting our results with the well-measured 
  semileptonic branching ratio of $B$ mesons we conclude that 
  the light sbottom interpretation of the experimental anomalies 
  is ruled out, unless $m_b \leq \msb{} + \msn{}$. \\[4mm] 
\end{abstract}

\vspace*{5mm}


\section{Introduction}

Supersymmetry predicts the existence of scalar partners for all
Standard Model fermions. Scalar quarks are usually assumed to be
heavy, based on direct searches at LEP~\cite{lep_ex,lep_theo} and the
Tevatron~\cite{tev_ex,tev_theo}.  However, most of the collider
searches rely on a large missing (transverse) energy cut, and
supersymmetric particles with small masses may escape detection,
because they lead to softer events with too little missing energy.  On
the other hand, decays of heavy Standard Model particles provide a
powerful tool to search for such light superpartners. No new particles
have been found in $Z$ decays at LEP-I and SLD. Hence supersymmetric
scenarios with particle masses below $m_Z/2$ are constrained, as their
couplings to the $Z$ boson must be very small.  Recently the ALEPH
collaboration has reported experimental hints for a light sbottom
squark with a mass around 4 GeV and a lifetime of 1 ps
\cite{aleph}. Its experimentally detected decay mode appears as a
chargino-mediated decay into a charm quark, a lepton and an
essentially massless anti-sneutrino \cite{aleph}. These findings have
prompted a reanalysis of an old anomaly in the MARK-I data for the
cross section of $e^+ e^- \to \textit{hadrons}$: the existence of a
squark with a mass between 3.6 and 3.7 \gev\ is found to bring the
measured cross section into agreement with the theoretical prediction
\cite{ps}.  Since the coupling of the $Z$ boson to sneutrino mass
eigenstates is constrained by the measured invisible $Z$ width, one
must supplement the Minimal Supersymmetric Standard Model (MSSM) by
right-handed neutrino and sneutrino states. The light sneutrino is
then predominantly right-handed. Interestingly, this model is
consistent with electroweak precision data and LEP limits on the mass
of the lightest CP-even Higgs boson \cite{chww}.\medskip

Yet the existence of a bottom squark $\tilde{b}$ with a mass
\emph{below} the $b$-quark mass and the conjectured chargino-mediated
semileptonic decay has a striking consequence: In such a scenario the
neutralino-mediated decays $b \to \tilde{b} \nu \tilde{\nu}{}^* $ and
$b \to \tilde{b} \ov{\nu} \tilde{\nu} $ are kinematically allowed.  In
the Standard Model, bottom quarks dominantly decay as $b \to c X$.
Hence $b$ decays are suppressed by the small element $V_{cb}\approx
0.04$ of the Cabibbo-Kobayashi-Maskawa (CKM) matrix. The new decay
modes do not suffer from any CKM suppression and therefore have
potentially large branching ratios. Since the light $\tilde{b}$ decays
only semileptonically, the supersymmetric decay channels increase the
lepton yield in $b$ decays through the cascade decay $ b \to \tilde{b}
\met \to c \ell \met$. Here $\met$ denotes the missing energy from the
(s)neutrinos. The measurement of the semileptonic branching ratio
$B_{SL}=Br( B \to X_{\ov{c}} \ell \ov{\nu}_{\ell})$, $\ell=e,\mu$, of
$B$ mesons is a mature field.  At the $B$ factory CLEO $B$,$\ov{B}$
pairs are copiously produced from the $\Upsilon (4S)$ resonance. The
measurement of $B_{SL}$ amounts to counting the leptons in the final
states of $B$-decays. The presence of secondary leptons from
non-leptonic $B$ decays followed by a semileptonic decay of the decay
products constitutes a background, which must be subtracted.  The
dilepton analysis by CLEO \cite{cleo} subtracts this background using
\emph{measured}\ branching ratios and lepton spectra. Hence it is
clear that the cascade decay $ b \to \tilde{b} \met \to c \ell \met$
would be ascribed to the signal rather than the background. The extra
events also pass the low cut $p_{\ell} \geq 600 \mev$ on the lepton
momentum in the $B$ rest frame, although leptons from the
supersymmetric cascade decays are softer than the primary leptons.
The CLEO result $B_{SL}= (10.49 \pm 0.46)\% $ \cite{cleo} and the LEP
measurement of $Br (b\to c \ell \ov{\nu}_{\ell}) =(11.01\pm 0.38 )\% $
\cite{blep} are consistent with the Standard Model prediction of $9.9
\% \leq B_{SL} \leq 13\% $ \cite{bbbg,ns}. $B_{SL}$ is the ratio of
the semileptonic and the total rate. In the Standard Model the CKM
element $V_{cb}$ drops out from this ratio and $B_{SL}$ depends only
on Standard Model parameters whose values are unaffected by a light
$\tilde{b}$. \bigskip

In this letter we investigate the contributions to $B_{SL}$ from $b
\to \tilde{b} [\to c \ell \tilde{\nu}{}^*] \nu \tilde{\nu}{}^* $ and
$b \to \tilde{b} [\to c \ell \tilde{\nu}{}^*] \ov{\nu} \tilde{\nu} $.
We calculate the rates of these bottom decays and the decay rate for
$\tilde{b} \to c \ell \tilde{\nu}{}^*$ in sect.~\ref{sect:l}. In
sect.~\ref{sect:p} we correlate the $\tilde{b}$ lifetime with the new
contribution to $B_{SL}$. We scan over the values of the
supersymmetric parameters entering the considered decay rates allowing
for a non-vanishing sneutrino mass. Finally we conclude in
sect.~\ref{sect:c}.

\section{Decay rates}\label{sect:l}
As mentioned in the Introduction, we have to extend the MSSM by
right-handed (s)neutrino states, because a purely left-handed light
sneutrino would couple to the $Z$ and would therefore contribute to
the well-measured invisible $Z$ width.\footnote{Still such a scenario
with dominantly right-handed sneutrinos might contradict cosmological
bounds \cite{toby} and one could need additional small R-parity
violating couplings to allow the LSP sneutrino to decay.}
Phenomenological constraints from flavor-changing neutral currents
further imply that the CKM matrix accompanying quark-squark-chargino
vertices is the same as in the couplings of quarks to the $W$ boson
\cite{ggms}. Hence the semileptonic $\tilde{b}$ decay is governed by
the same CKM element $V_{cb}\approx 0.04$ as the standard semileptonic
$b$ decay. We further assume that there are two light sneutrino states
corresponding to $\tilde{\nu}_{e}$ and $\tilde{\nu}_{\mu}$, so that
both semileptonic decays with $\ell=e$ and $\ell =\mu$ are possible.
This assumption, however, does not influence the correlation between
the sbottom lifetime and the branching ratio $Br ( b \to \tilde{b}
\met)$, because it amounts to an overall factor of 2 for all relevant
decay rates. We calculate all decay rates at the partonic level, in
the tree level approximation of perturbation theory.  Although $b$ and
$\tilde{b}$ hadronize, the binding effects are suppressed by two
powers of $\Lambda_{QCD}/E$, where $E$ is the average energy release
to the final state hadron \cite{hqe}.  We account for these power
corrections and for contributions of uncalculated radiative
corrections by conservatively inflating the allowed ranges for the
input parameters in our phenomenological discussion in
sect.~\ref{sect:p}. Further, we remark that our formulae become
inaccurate for sneutrino masses near the kinematic limit. In this case
$E= {\cal O}(\Lambda_{QCD})$, and the final state hadron moves too
slowly in the rest frame of the decaying hadron. Therefore naive
perturbation theory breaks down. Yet we will see in sect.~\ref{sect:p}
that in this region it is hard to accommodate for the conjectured
sbottom lifetime.\medskip

We denote the light sbottom and light sneutrino mass eigenstates by
$\tilde{b}_1$ and $\tilde{\nu}_1$. The mixing angle $\theta_b$
relating $\tilde{b}_1$ to scalar partners of the chiral $b$-fields is
defined as $\tilde b_{1} = \cos \theta_b \tilde b_{\rm L} + \sin
\theta_b \tilde b_{\rm R} $, with an analogous definition for the
sneutrino mixing angle $\theta_{\nu}$. 
We adopt the standard notation \cite{haka} for the MSSM mass
parameters and mixing matrices: the chargino mass matrix is
diagonalized as
\begin{equation}
{\cal M}_c \; = \;  \lt( 
\begin{array}{cc}
M_2                      & \sqrt{2} m_W \sin \beta \\
\sqrt{2} m_W \cos \beta  & \mu 
\end{array}
\rt) 
\; = \; 
U^T \, \mbox{diag} \lt( \mch{j} \rt) 
        \, V .  \label{chm}
\end{equation}
By convention $M_2 > 0$ and $\cp{1}$ is the lighter chargino. 
Since we are not interested in CP violation here, we choose all
mass matrices and the unitary mixing matrices $U$ and $V$ real.  
$\tan \beta$ is the ratio of the Higgs vacuum expectation values. 
The neutralino mass matrix reads
\begin{eqnarray}
 \lt( 
\begin{array}{r@{\hspace{2ex}}r@{\hspace{2ex}}r@{\hspace{2ex}}r}
M_1      &      0 &  - \, m_Z  s_w \cos \beta
                  &     m_Z  s_w \sin \beta \\
0       &    M_2 &   m_Z  c_w \cos \beta    
                 &   - m_Z c_w \sin \beta  \\
 - m_Z s_w \cos \beta & 
 m_Z c_w \cos \beta &   0   &       - \mu \\
 m_Z s_w \sin \beta & 
- m_Z c_w \sin \beta &     - \mu     &    0   
\end{array}
\rt)  
\; =
N^T \,  \mbox{diag} \lt( \mne{j} \rt) 
        \, N \, , \label{nem}    
\end{eqnarray}
where $s_w, c_w$ are the sine and cosine of the weak mixing angle.
We choose $N$ as an orthogonal matrix. Then the mass eigenvalues can
be negative and the physical neutralino masses are their absolute
values.\medskip

\begin{figure}[t] 
\begin{center}
\includegraphics[width=12.0cm]{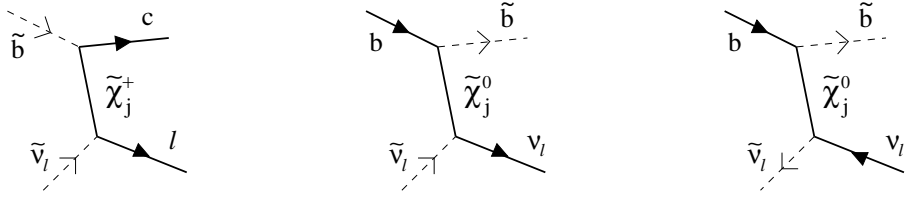}
\end{center} \vspace*{0mm}
\caption[]{\label{fg:diags} {\sl Semileptonic sbottom decay (left) and the decays 
    $b \to \tilde{b} \nu_l \tilde{\nu}_l^* $ (center) and $b \to
    \tilde{b} \ov{\nu}_l \tilde{\nu}_l $ (right).  $l$ represents $e$
    or $\mu$.}}
\end{figure}

The decay $\tilde{b}_1 \to c \ell \tilde{\nu}_1^*$ is depicted in the
left diagram of Fig.~\ref{fg:diags}. 
The corresponding matrix element equals $K \, \ov{c} P_R \ell$, where
$P_R=(1+\gamma_5)/2$ and 
\begin{eqnarray}
K &=& V_{cb} \, g \, \cos \theta_{\nu}  \, \sum_{j=1}^2 
         \frac{V_{j1}}{m_{\chi_j}} \, 
        \lt[ g\, \cos \theta_b \, U_{j1} \, - \, 
             h_b \, \sin \theta_b \, U_{j2} \rt] . \label{defk}   
\end{eqnarray}
Here we have neglected the sbottom momentum in the chargino propagator
and the small Yukawa coupling to $\ell$. $g\simeq 0.65 $ is the SU(2)
gauge coupling and $h_b$ is the bottom Yukawa coupling.  Our result
agrees with the expression in \cite{hk}.  We conveniently re-express
$V_{jk}$, $U_{jk}$ and $\mch{j}$ in terms of $M_2$, $\mu$ and $\beta$:
\begin{eqnarray}
\sum_{j=1} V_{ji} \, \frac{1}{m_{\chi_j}} \, U_{jk} &=& 
        \lt[ V^T \, D^{-1} \, U \rt]_{ik} \; = \; \lt[ {\cal M}_c^{-1} \rt]_{ik}
        . \label{cinv} 
\end{eqnarray}
Then \eq{defk} becomes 
\begin{eqnarray}
K &=& V_{cb} \, g \, \cos \theta_{\nu}  \, 
        \frac{g \, \mu \, \cos \theta_b \, + \, \sqrt{2} \, h_b \, m_W
        \, \sin \theta_b \, 
        \sin \beta}{ M_2 \, \mu - m_W^2 \, \sin (2 \beta) } 
        . \label{kfine}
\end{eqnarray}
A lifetime of ${\cal O}$(1 ps)~\cite{aleph} implies that the
semileptonic decay rate of the light sbottom ($\propto
m_{\tilde{b}_1}^3/\mch{1}^2$) exceeds the rate of the $b$ quark
($\propto m_b^5/m_W^4$) by roughly a factor of 7. To accommodate for
this, one is naturally lead to the portion of the parameter space with
a large Yukawa coupling $h_b$ and thereby a large $\tan \beta$. In
this region supersymmetric QCD corrections to the relation of $h_b$ to
$m_b$ can be huge \cite{copw}.  Yet such $\tan \beta$ enhanced
corrections can be summed to all orders \cite{cgnw}:
\begin{eqnarray}
h_b &=& \frac{g}{\sqrt{2} m_W \cos \beta } 
     \; \frac{ m_b }{1+\Delta m_b^{SQCD}} 
\label{hbsum}
\end{eqnarray}
with 
\begin{eqnarray}
\Delta m_b^{SQCD}  & = &  \frac{2\alpha_s}{3\pi}\, \mg 
                        \mu\, \tan \beta \,I(\msb{1},\msb{2},\mg)\,
        . \label{dmb} \nn 
 I(a,b,c) & = &
        \frac{1}{(a^2-b^2)(b^2-c^2)(a^2-c^2)} \, 
        \lt(   a^2 b^2 \ln \frac{a^2}{b^2}
              +b^2 c^2 \ln \frac{b^2}{c^2}
              +c^2 a^2 \ln \frac{c^2}{a^2}  \rt) . \no
\end{eqnarray}
Here $\mg$ is the gluino mass. The strong coupling constant $\alpha_s$
must be evaluated at a high scale of order $\mg$ or $\msb{2}$.
Moreover, the light sbottom has to be included in the running of
$\alpha_s$. The non-leading corrections to \eq{hbsum} can be safely
neglected~\cite{decay}.  The semileptonic decay rate now reads
\begin{eqnarray}
\Gamma_{SL} \; =\; 
        \Gamma \lt( \tilde b_1 \to c \ell \tilde \nu_2  \rt) 
&=& 
        \frac{|K|^2\, }{32 \msb{1} \pi^5} \, 
        p\lt( \msb{1} , m_c, m_{\tilde \nu} \rt) .\label{gsl}
\end{eqnarray}
Since we assume that the two decay channels with $\ell = e$ and $\ell
=\mu$ are allowed, the sbottom lifetime is given by $\tau_{\tilde
b}=1/\Gamma_{tot}=1/(2\, \Gamma_{SL})$. The phase space
integral reads
\begin{equation}
p \lt( M, m_1, m_2  \rt) = 
 \int \frac{d^3 \vec p_1 }{2 E_1}  \frac{d^3 \vec p_2 }{2 E_2} 
  \frac{d^3 \vec p_3 }{2 E_3} \, \delta^{(4)} \lt( P-p_1-p_2-p_3 \rt)
    \, p_1 \cdot p_3, 
\; = \; M^4 \, p \lt( 1, \frac{m_1}{M}, \frac{m_2}{M}  \rt)
\end{equation}
with the masses $p_3^2=0,~p_{1,2}^2=m_{1,2}^2$ and $P^2=M^2$. 
We find
\begin{eqnarray}
p \lt( 1, x_1, x_2  \rt) &=& 
        \frac{\pi^2}{8} \, \lt[ 
         x_1^4 \lt( 1 + x_2 \rt)^2 \, y \, \ln \frac{y+x}{y-x} + 
        \lt[ 4 x_1^2 x_2^2 - \lt( 1 + x_2^2 \rt) \lt( x_1^4 + 2 x_2^2
        \rt) \rt] \ln \frac{1+x}{1-x} \rt. \nn 
&& \phantom{\frac{\pi^2}{8} \,} 
   \lt. \, + \,  \frac{ x \, \lt[ \lt( 1+ x_2 \rt)^2 - x_1^2  \rt] }{6}
        \, \lt[ 1 + 10 x_2^2 + x_2^4 - 5 x_1^2 \lt( 1+x_2^2 \rt) 
                - 2\, x_1^4  \rt] \rt] \nn 
&& \qquad \qquad \qquad 
   \mbox{with}\quad x=\sqrt{\frac{(1-x_2)^2-x_1^2}{(1+x_2)^2-x_1^2}},
              \qquad y = \frac{1-x_2}{1+x_2}. 
\nn
p \lt( 1, x_1, 0  \rt) &=&  
   \frac{\pi^2}{48} \, \lt[ \lt(1-x_1^2)(1 -5 x_1^2 -2 x_1^4  \rt) 
                           - 12\, x_1^4 \, \ln x_1    \rt]  \no
\end{eqnarray}

Next we turn to the two neutralino-mediated decays $b \to \tilde{b}_1
\nu \tilde{\nu}_1^* $ and $b \to \tilde{b}_1 \ov{\nu} \tilde{\nu}_1 $
in Fig.~\ref{fg:diags}.  The two different final states are possible
because of the Majorana nature of the neutralinos. As for the sbottom
decay we assume two light flavor-generations of neutrinos.  The matrix
element for $b \to \tilde{b}_1 \nu \tilde{\nu}_1^*$ can be written as
$K_R \ov{\nu} P_R b$ and the amplitude for $b \to \tilde{b}_1 \ov{\nu}
\tilde{\nu}_1 $ is of the form $K_L \ov{\nu} P_L b$ with
$P_L=(1-\gamma_5)/2$. The coefficients are
\begin{eqnarray}
K_R &=&  g  \cos \theta_\nu \, \sum_{j=1}^4 \, 
        \frac{N_{j2} \, - \, N_{j1} t_w   }{m_{\chi_j^0}} \, 
        \lt[ \frac{g}{3} \, t_w \, \sin \theta_b \, N_{j1}
        \, + \, \frac{h_b}{\sqrt{2}}  \, \cos \theta_b
        \, N_{j3}  \rt] \nn[2mm] 
&=& - \frac{g \cos \theta_\nu}{d}  
        \lt[ g\, \mu^2 \, \frac{M_2 \, t_w^2}{3} \, \sin \theta_b 
         \, + \, \frac{h_b}{\sqrt{2}} \, 
         \mu  m_W \sin \beta \, \cos \theta_b \,
        \lt( M_1 +  M_2 t_w^2 \rt) \rt] 
        \label{krfine} \\[3mm]
K_L &=& g  \cos \theta_\nu \, \sum_{j=1}^4 \, 
        \frac{N_{j2} \, - \, N_{j1} t_w }{m_{\chi_j^0}} \, 
        \lt[ \frac{g}{2} \, \cos \theta_b  \, 
        \lt(   \frac{t_w}{3}  \, N_{j1} \, - \, N_{j2} \rt) 
        \, + \, \frac{h_b}{\sqrt{2}} \, \sin \theta_b
        \, N_{j3}  \rt] \nn[2mm]
&=& - \frac{g \cos \theta_\nu}{d}  
        \lt[  g\, \mu^2  
  \lt( \frac{M_2 \, t_w^2}{6} + \frac{M_1}{2} \rt) 
        \cos \theta_b 
         \, + \, \frac{h_b}{\sqrt{2}} \, 
         \mu  m_W \sin \beta \, \sin \theta_b \,
        \lt( M_1 +  M_2 t_w^2 \rt) \rt] 
        \label{klfine} . 
\end{eqnarray}
$t_w$ is given by $\tan \theta_w=s_w/c_w$, and  
$-d$ is the determinant of
the neutralino mass matrix in \eq{nem}:
\begin{eqnarray}
d & =& \mu \, \lt[ M_1 \, M_2 \, \mu - m_W^2 \, \sin (2 \beta) \,
        \lt( M_1 +  M_2 t_w^2 \rt) \rt] . \no
\end{eqnarray} 
The couplings given in \eq{krfine} and \eq{klfine} nicely reveal that
the term proportional to $h_b$ is zero for $M_1=-M_2 t_w^2$. In this
case the $\tilde{Z} \tilde{H}_1$ element of the inverse neutralino
mass matrix vanishes. Then the Higgsino $\tilde{H}_1$, which couples
with $h_b$ to the $b$ and $\tilde{b}$, cannot propagate into the Zino,
which is the only gauge fermion coupling to the $\nu$-$\tilde{\nu}$
line.  By comparing \eq{krfine} and \eq{klfine} with \eq{kfine} one
can also identify the terms in the sbottom and bottom decay amplitudes
which are related by electroweak SU(2) symmetry.  The SU(2) symmetry
leads to a high correlation between the two decay modes: if one picks
SUSY parameters keeping $K$ large enough to accommodate the lifetime
observed in \cite{aleph}, one cannot simultaneously make both $K_{R}$
and $K_{L}$ arbitrarily small.

After summing the decay rates for the two decay modes, 
each lepton flavor $\ell =e,\mu$ gives a new contribution to the
bottom width of 
\begin{eqnarray}
\Delta \Gamma &\equiv & 
\Gamma \lt( b \to \tilde{b}_1 ( \nu \tilde{\nu}_1^* + 
                                \ov{\nu} \tilde{\nu}_1) \rt)
\; = \; 
        \frac{|K_R|^2+|K_L|^2}{64 m_b \pi^5} \, 
        q \lt( m_b, \msb{1}, \msn{1} \rt) 
        . \label{gb}
\end{eqnarray}
Here the space space integral reads 
\begin{equation}
q \lt( M, m_1, m_2  \rt) = 
 \int \frac{d^3 \vec p_1 }{2 E_1}  \frac{d^3 \vec p_2 }{2 E_2} 
  \frac{d^3 \vec p_3 }{2 E_3} \, \delta^{(4)} \lt( P-p_1-p_2-p_3 \rt)
    \, P \cdot p_3  
\; = \; M^4 \, q \lt( 1, x_1, x_2  \rt),
\end{equation}
where again $p_3^2=0,~p_{1,2}^2=m_{1,2}^2$ and $P^2=M^2$. We calculate
\begin{eqnarray}
q \lt( 1, x_1, x_2  \rt) &=& 
        \frac{\pi^2}{8} \, \lt[ 
         \lt( x_1 - x_2 \rt)^2 \, y \, \ln \frac{y+x}{y-x} \, - \, 
        \lt[ x_1^2 \, + \,  x_2^2 \, - \, 4\,  x_1^2 x_2^2 \, + \,  2\,
        x_1^4 x_2^2 \, + \,  2\, x_1^2 
             x_2^4 \, \rt] 
          \ln \frac{1+x}{1-x} \rt. \nn 
&& \phantom{\frac{\pi^2}{8} \,} 
   \lt. \, + \,  \frac{ x \, \lt[ 1 - \lt( x_1 - x_2 \rt)^2 \rt] }{6}
        \, \lt[ 2\, + \, 5\, x_1^2 + 5\, x_2^2 \, - \, x_1^4  \, - \, x_2^4 
                \, - \, 10\, x_1^2 x_2^2 \, - \, x_2^4 
                \rt] 
        \rt] \nn 
&& \qquad \qquad \qquad 
   \mbox{with}\quad x=\sqrt{\frac{1-(x_1+x_2)^2}{1- (x_1-x_2)^2 }},
              \qquad      y = \frac{x_1 + x_2}{x_1 - x_2}. 
\nn
q \lt( 1, x_1, 0  \rt) &=&  
   \frac{\pi^2}{48} \, \lt[ \lt(1-x_1^2)(2 +5\, x_1^2 -x_1^4  \rt) 
                           \, + \, 12\, x_1^2 \ln x_1    \rt] . \no
\end{eqnarray}
In a $B$-factory like CLEO the $B$ mesons move too slowly to resolve a
displaced vertex from the $\tilde{b}_1$ in the cascade decay $ b \to
\tilde{b}_1 \met \to c \ell \met$. Hence the signal of a $b$ decay
into a light sbottom would basically be an increase of the
semileptonic branching ratio $B_{SL}$ and a shift of the lepton spectrum
to lower energies as discussed in the Introduction.  Therefore it is
useful to normalize $\Delta \Gamma$ to the semileptonic decay rate
$\Gamma_{SL}$ of the $B$ meson (which to order $\Lambda_{QCD}^2/m_b^2$
coincides with the semileptonic decay rate of the $b$ quark
\cite{hqe}).  The experimental value for $\Gamma_{SL}$ is obtained by
dividing the measured $B_{SL}=(10.49 \pm 0.46)\%$ \cite{cleo} by the
measured lifetime $\tau_{B}=1.55\pm0.03\,$ps${}^{-1}$ \cite{pdg}:
\begin{eqnarray}
    \Gamma_{SL} & = & \Gamma ( B \to X_{\ov{c}} \ell^+ \nu_{\ell}  ) 
        \; = \; (4.45 \pm 0.21) \cdot 10^{-14} \gev 
        \no
\end{eqnarray}
per lepton flavor $\ell=e$ or $\mu$.  SUSY parameters yielding $\Delta
\Gamma > 5$ are already excluded from the measured $B$ lifetime alone:
since the total branching ratio into light leptons is $2 B_{SL}\simeq
20\% $, $\Delta \Gamma $ would exceed the total decay rate
$1/\tau_{B}$ in these scenarios. We further remark that the decay mode
$ b \to \tilde{b}_1 \met \to c \ell \met$ would also influence the
determination of $V_{cb}$, which enters our analysis of the sbottom
lifetime, from inclusive semileptonic decays. The true value of
$V_{cb}$ would be somewhat lower than the Standard Model value of 0.04
and our exclusion plots in the following section would become even
more restrictive. On the other hand, measurements of $V_{cb}$ from
exclusive decays near the kinematic endpoint are less affected because
of the softer leptons from the supersymmetric decays.

\section{MSSM Parameter Space}\label{sect:p}

\begin{figure}[b] 
\begin{center} \vspace*{-4mm}
\includegraphics[width=15.0cm]{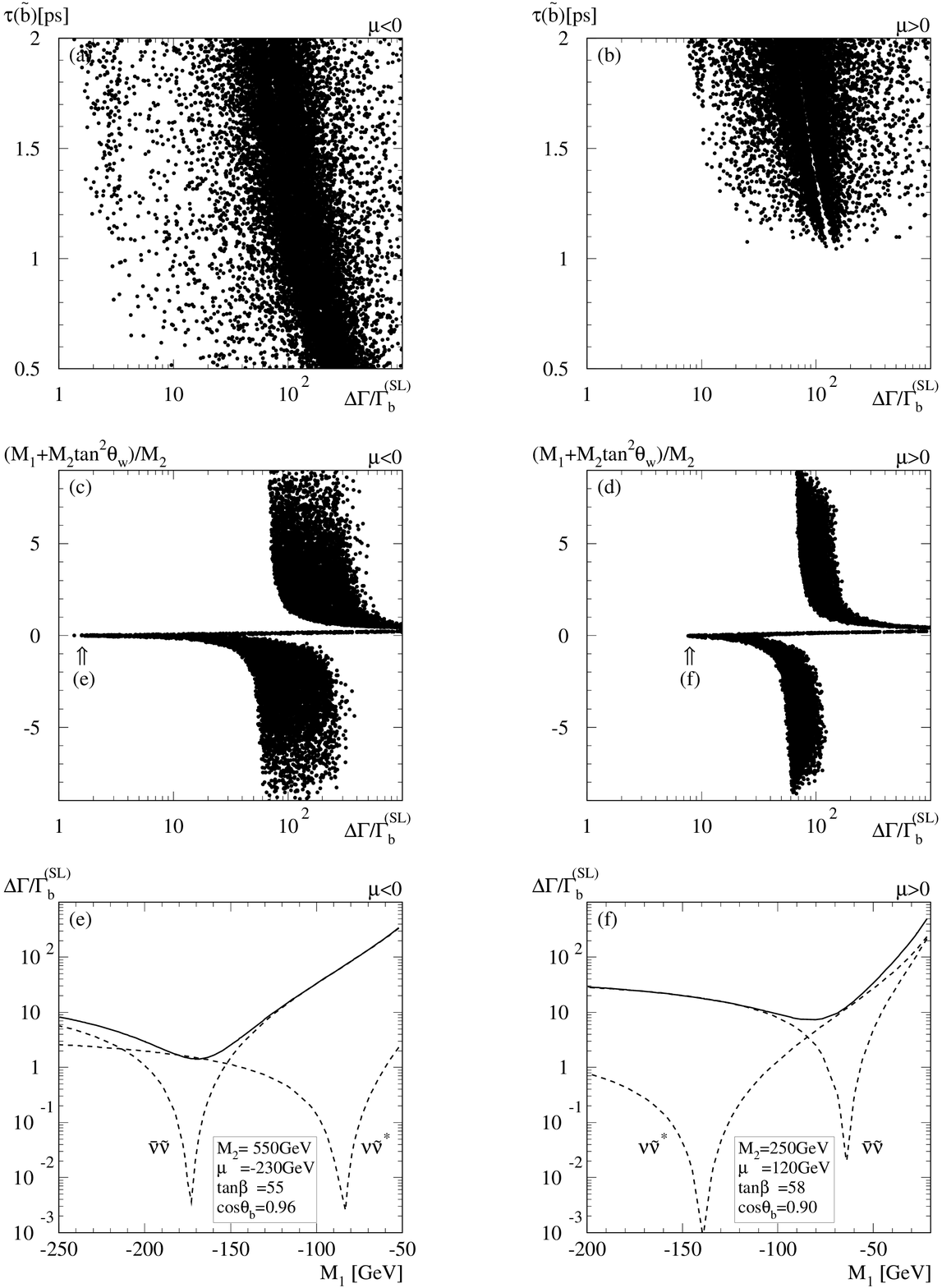}
\end{center} \vspace*{-4mm}
\caption[]{\label{fg:all} {\sl
    The contribution of $b \to \tilde{b} + \met \to c \ell \met$ 
    to the semileptonic $b$ decay in
    20000 MSSM scenarios, all fulfilling the relaxed sbottom lifetime
    requirement of $\tau_{\tilde{b}}<2$~ps 
    (left: $\mu<0$, right: $\mu>0$.  The SUSY
    bottom decay width is plotted versus the sbottom lifetime (a,b) and
    versus the relevant combination of gaugino masses (c,d). For two
    particular parameter points the variation of the different decay
    channels with $M_1$ is shown (e,f).}}
\end{figure}

To determine how the semileptonic sbottom and the bottom decay widths
are related, we perform an MSSM parameter scan: for all models leading
to a sbottom lifetime between 0.5 and 2~ps we compute the additional
semileptonic bottom decay width $\Delta \Gamma$ and compare it to the
measured value, as shown in Fig.~\ref{fg:all}(a,b). In the scan we
assume a sbottom mass of $3.5 \gev$ and fix the presumably
small~\cite{aleph} sneutrino LSP mass to $0.3 \gev$.  Disregarding the
large theoretical errors we emphasize that the results of the analysis
become dependent on the sneutrino mass only close to threshold. For
reasons discussed below, the case $\mu>0$ hardly ever leads to a
sbottom lifetime below 2~ps, whereas $\mu<0$ generates a rich variety of scenarios.
In particular, the case $\mu>0$ cannot accommodate sneutrino masses
above $0.5 \gev$. \smallskip

We note that all Yukawa couplings contributing to the considered
decays are enhanced by $\tan \beta$. \ie to reach the measured sbottom
lifetime one is automatically driven into the large $\tan \beta$
regime. We in fact observe that varying all other input parameters in
the given ranges only allows values of $\tan \beta>15$ for $\mu<0$ and
$\tan \beta>25$ for $\mu>0$. As an upper limit we choose $\tan \beta
=60$. Since both the bottom and the sbottom decay widths are enhanced,
the minimal $\Delta \Gamma$ in the scan depends only weakly on the
value of $\tan \beta$ in the given interval.

The parameters determining the masses of the virtual charginos and
neutralinos are $M_2$ and $\mu$. In addition the Higgsino mass
parameter enters into the correction to the bottom Yukawa coupling:
for positive values of $\mu$ the mass correction $\Delta m_b$ is
positive, leading to a decrease in $h_b$ and therefore a decrease in
the sbottom decay width until the gaugino coupling to the sbottom
becomes dominant, at the expense of the total decay width. For $\mu<0$
the mass correction becomes negative. Values around $\Delta m_b = -1$
dramatically increase the Yukawa coupling\footnote{Models with
  $h_b^2/(4 \pi^2)>1$ we reject as non-perturbative.}. A comparison of
scenarios with the two signs of $\mu$ in Fig.~\ref{fg:all}(a,b) shows
that the impact of the increased Yukawa coupling leads to an
enhancement by a factor of two in the typical sbottom lifetime.\smallskip

Since the sbottom decays through a virtual chargino, either $M_2$ or
$|\mu|$ has to be sufficiently small, to keep the suppression
moderate. For $\mu>0$ both parameters $|\mu|$ and $M_2$ have to to be
smaller than $400 \gev$. The light chargino mass, which we require to
respect the LEP lower limit of $103 \gev$\footnote{This limit is based
on a neutralino LSP scenario, but the sneutrino LSP does hardly change
the signature and leads to an increased production cross section. We
therefore assume a chargino mass limit close to the kinematical
limit~\cite{lepc}.}, is found to be smaller than $140 \gev$.  In the
less constrained case of $\mu<0$ only $M_2$ has a strict upper limit
of $500 \gev$, but large values of $|\mu| \gtrsim 500 \gev$ require
$M_2 \lesssim 250 \gev$.  The upper limit for the mass of the light
chargino becomes $350 \gev$. We vary the additional Bino mass
parameter $M_1$ between $\mp 1 \tev$ to always cover the decoupling
point $M_1 = -M_2 t_w^2$, as described in sect.\ref{sect:l}. In
Fig.\ref{fg:all}(c,d) we show that $\Delta \Gamma< 50 \Gamma_{SL}$ can
only be achieved for parameters close to this decoupling point.  In
particular for the Yukawa coupling dominated models with $\mu <0$ we
observe a sharp decrease in the minimum value for $\Delta \Gamma$.
The numerical width of the allowed parameter region is shown in
Fig.~\ref{fg:all}(e,f). With $\cos \theta_b \sim 1.0$ the $\bar{\nu}
\tilde{\nu}_1$ decay channel dominates. For $M_1=-M_2 t_w^2$ the
Yukawa contribution vanishes, and the corresponding gauge coupling
proportional to $\sin \theta_b$ leads to negligible values of $\Delta
\Gamma$. In the other decay channel $\nu \tilde{\nu}_1^*$ the gauge
coupling is enhanced by $\cos \theta_b$ and thereby rescues the total
MSSM contribution to the semileptonic decay width. However, typical
values of $\Delta \Gamma$ become significantly smaller, in particular
for $\mu<0$, where the large bottom Yukawa coupling was further
enhanced. By contrast, models sufficiently separated from the
decoupling point easily yield an enhancement of several hundred times
the Standard Model value of $\Gamma_{SL}$.\medskip

The sbottom and the sneutrino mixing angle are constrained by the
measurement of the $Z$ width: both particles have to decouple from the
$Z$ boson. A right-handed LSP sneutrino does indeed not couple to the
$Z$. However, it does not couple to the intermediate chargino in the
sbottom decay either. We therefore assume a fraction of
left-handedness in the LSP, parameterized by $\cos \theta_\nu<0.2$. A
small fraction of left-handedness might be a hint for a see-saw
mechanism in the scalar neutrino mass matrix. Since both the bottom
and the sbottom decay width scale with the square of this fraction, we
have checked that reducing $\cos \theta_{\nu}$ does not affect the
result, until it suppresses the sbottom decay widths too strongly to
allow for any models with a lifetime $\tau_{\tilde{b}}<2$~ps.

The light sbottom decouples from the $Z$ for a leading order mixing
angle of $|\cos \theta_b| = s_w \sqrt{2/3}$, \ie a mixture of left and
right-handed states aligned with the weak mixing angle. Taking into
account possible experimental uncertainties we impose $0.8<|\sin
\theta_b|<1.0$. Since the Yukawa coupling to the chargino is dominant
in most of our valid models, the mixing angle strongly affects the
sbottom decay width. A completely right-handed sbottom is preferred,
because it gives the Yukawa coupling a maximal relative weight.
However, the Majorana nature of the neutralino allows the two decay
modes in Fig.~\ref{fg:diags}, which couple to either the right-handed
or the left-handed sbottom states. Therefore $\cos \theta_b=1$ also
leads to an enhancement of the respective decay channel, while
suppressing the other. Even at the decoupling point $M_1=-M_2 t_w^2$
it is impossible to switch off both decays simultaneously.

\section{Conclusions}\label{sect:c}
We have investigated implications of recently reported experimental
hints for a light sbottom squark $\tilde{b}$ with a mass below $m_b$
and a lifetime around 1 picosecond decaying as $\tilde{b} \to c \ell
\tilde{\nu}^*$. We have studied the decay modes $b \to \tilde{b} \nu
\tilde{\nu}^*$ and $b \to \tilde{b} \ov{\nu} \tilde{\nu} $, which are
related to the semileptonic sbottom decay by electroweak SU(2)
symmetry. At $B$-factories these decay modes would manifest themselves
through the cascade decay $ b \to \tilde{b} \met \to c \ell \met$ and
would increase the well-measured semileptonic branching ratio $B_{SL}$
of $B$ mesons. We have determined the correlation between the sbottom
lifetime and the rates of these supersymmetric $b$ decays.  A scan
over the entire MSSM parameter space has shown that the rate of $b \to
\tilde{b} \met $ typically exceeds the semileptonic bottom decay rate
$\Gamma_{SL}$.  It easily reaches values which are up to 1000 times
the experimental value $\Gamma_{SL}=B_{SL}/\tau_B$. The minimal value
is $\Gamma (b \to \tilde{b} \met) \approx 1.4 \Gamma_{SL}$ for $\mu<0$
and $\Gamma (b \to \tilde{b} \met) \approx 7 \Gamma_{SL}$ for $\mu>0$.
Both are obtained for large values of the bottom Yukawa coupling. The
minimal values of $\Gamma (b \to \tilde{b} \met)$ correspond to a
small region of the supersymmetric parameter space in which a
Zino-Higgsino mixing term in the neutralino sector vanishes.
In view of the the good agreement of the measured $B_{SL}$ with the
Standard Model prediction we conclude that experimental anomalies
reported in \cite{aleph,ps} cannot be interpreted as light sbottoms
decaying as $\tilde{b} \to c \ell \tilde{\nu}^*$, unless the decays $b
\to \tilde{b} \nu \tilde{\nu}^*$ and $b \to \tilde{b} \ov{\nu}
\tilde{\nu} $ are kinematically forbidden.

We remark here that our reasoning similarly constrains a light sbottom
interpretation of the anomalies of \cite{aleph,ps}, if the
sbottom is \emph{heavier} than the bottom quark. If the decays
$\tilde{b} \to b \nu \tilde{\nu}^*$ and $ \tilde{b} \to b \ov{\nu}
\tilde{\nu} $ are kinematically allowed, they will by far be the
dominant decay modes in most of the supersymmetric parameter space and
the observed decay $\tilde{b} \to c \ell \tilde{\nu}^*$ would be
rare. This would point at a much higher sbottom production rate and we
presume that the secondary vertices from the $\tilde{b}$ and $b$ in
these $\tilde{b} \to b \met$ decays would have been detected in
collider experiments. While a detailed study of this scenario is
beyond the scope of this letter, we also consider this possibility as
remote. The anomalies reported in \cite{aleph} are essentially only 
compatible with a light sbottom interpretation, if $|m_b-\msb{1}| \leq
\msn{1}$. 

Even if we leave the framework of supersymmetry, it is hard to relate
the experimental anomalies to some other bottom-flavored
object. Consider any new SU(2)-invariant renormalizable model with
conserved lepton number: the semileptonic decay mode will then have
the topology of the left diagram in Fig.~\ref{fg:diags}. By SU(2)
symmetry then $b$ decays corresponding to the middle diagram are
allowed. The suppression of this decay mode would involve fine tuning
of the left and right-handed $b$ flavor components and between the
U(1) and SU(2) gauge sectors.

\section*{Acknowledgments}
U.N.\ thanks Marcela Carena, Carlos Wagner and Frank W\"urthwein for
stimulating discussions. T.P.\ wants to thank Toby Falk, Tao Han and
Stephan Lammel for useful discussions and Toby Falk for carefully
reading the manuscript.

\end{document}